# Personnalisation de bases de données multidimensionnelles

**F. Ravat, O. Teste, G. Zurfluh**

*IRIT (SIG/ED)*
*118, route de Narbonne*
*31062 Toulouse cedex 04*

*{ravat, teste, zurfluh}@irit.fr*

*RÉSUMÉ. Nos travaux se situent dans le cadre des systèmes décisionnels reposant sur une modélisation multidimensionnelle des données. Plus précisément, notre objectif est de proposer un ensemble de concepts et de mécanismes pour la spécification de bases de données multidimensionnelles personnalisées. Cette personnalisation consiste en l'association de poids aux différents composants d'un schéma multidimensionnel. La définition de cette personnalisation s'effectue à l'aide d'un langage reposant sur le principe Evènement Condition Action. Cette personnalisation impacte aussi bien l'affichage des données décisionnelles que leurs analyses (au travers des opérations de forage, rotation…)*

*ABSTRACT. This paper deals with decision support systems resting on multidimensional modelling of data. Moreover, we intend to offer a set of concepts and mechanisms for personalized multidimensional database specifications. This personalization consists in associating weights to different components of a multidimensional schema. Personalization specifications are specified through the use of a language based on the principle of Event Condition Action. This personalisation determines multidimensional data display as well as their analyses (with the use of drilling or rotating operations).*

*MOTS-CLÉS : modélisation et analyses multidimensionnelles, personnalisation, règles ECA*

# 1. Introduction

Les **systèmes décisionnels** visent à transformer les données opérationnelles en informations pouvant être interprétées par les décideurs. Ils reposent souvent sur des espaces de stockages dédiés appelés entrepôts et magasins de données (« *data warehouses* », « *data marts* »). L'entrepôt de données centralise les données décisionnelles de manière uniformisée tout en assurant la cohérence, la persistance et l'historisation de ces données. Les magasins de données visent à supporter efficacement les analyses décisionnelles et sont généralement dédiés à une classe de décideurs et/ou un thème d'analyse. L'organisation des données au sein d'un magasin suit une modélisation multidimensionnelle ; on parle de **Base de Données Multidimensionnelles (BDM)**. Cet article se situe dans le cadre des magasins de données multidimensionnelles.

## 1.1. Problématique

La mondialisation et la concurrence exacerbée réclament de la part des décideurs une grande réactivité. Ces décideurs exigent des outils suffisamment flexibles et évolutifs pour faciliter leurs tâches quotidiennes d'analyses. A l'heure actuelle, la diversité des besoins d'analyse est satisfaite dans les systèmes décisionnels classiques par la mise en place de divers magasins orientés sujets. Or, la conception d'une BDM est une tâche complexe et le plus souvent assez longue qui à l'heure actuelle ne repose pas sur des concepts, des formalismes, une démarche et une méthode standards (Rizzi et al., 2006). De plus, l'implantation d'un système décisionnel composé de nombreux magasins de données multidimensionnelles nécessite la mise en place de processus d'élaboration, d'alimentation et de rafraîchissement des données lourds sans oublier des efforts de maintenance importants. Il est alors difficile d'envisager un magasin de données multidimensionnelles pour chaque décideur. Les systèmes actuels s'avèrent imparfaits, voire inadaptés à ces exigences d'**adaptation**.

**Notre problématique consiste à proposer un système décisionnel personnalisable pour chaque décideur.** En fait, nous souhaitons proposer un modèle de BDM configurable pour répondre aux besoins des décideurs.

## 1.2. Travaux existants

L'idée de développer des mécanismes permettant de personnaliser un système informatique n'est pas une idée nouvelle. Notamment, nous pouvons citer les travaux du domaine de la recherche d'information (RI) (Korfhage, 1997) (Bouzeghoub, Kostadinov, 2005). La personnalisation en RI consiste généralement à définir des profils utilisateurs (ensembles plus ou moins structurés de caractéristiques) qui sont exploités aux différentes étapes du processus de RI : indexation, recherche,…

La personnalisation de BD décisionnelles a fait l'objet d'une première proposition (Bellatreche et al., 2005) visant à fournir aux décideurs la visualisation la plus adaptée à ces requêtes. Pour ce faire, les auteurs proposent des contraintes de visualisation précisant la structure du cube de données résultat et des préférences utilisateurs définies à l'aide du concept de pré-ordre total (Koutriba, Ionnidis, 2004) portant sur un attribut d'une dimension. Cette proposition se limite à des cubes de données dont les dimensions contiennent un seul attribut et à l'opération de sélection des données pour calculer le cube résultat.

D'autres travaux ont développé des entrepôts de données actives (« Active Data Warehouses ») (Thalhammer, et al., 2001), (Thalhammer, Schrefl, 2002). La proposition consiste à automatiser les tâches d'analyse récurrentes (par exemple, édition hebdomadaire d'un même tableau de bord). Contrairement à notre objectif, cette proposition se focalise sur les aspects internes d'organisation des données et d'optimisation des calculs sur les données.

### 1.3. Contributions et organisation

Notre objectif est de proposer pour chaque décideur un environnement personnalisé reposant sur un modèle de données multidimensionnelles. Cette personnalisation permet de mettre en valeur les données essentielles tout en facilitant leur manipulation lors des analyses décisionnelles. Pour ce faire, nous souhaitons proposer une solution plus complète que celle de (Bellatreche et al., 2005). Cette personnalisation va permettre d'associer un poids ou une préférence à tout composant d'une BDM voire aux valeurs de ces composants. Nous souhaitons également proposer un langage de personnalisation reposant sur le concept des règles actives. Cette personnalisation doit impacter aussi bien l'affichage final des données décisionnelles que leurs analyses au travers des opérations de forages et de rotation de cubes de données (changement d'axes d'analyse).

Cet article repose sur le plan suivant. En section 2, nous présentons notre modèle de données multidimensionnelles. Cette section intègre aussi bien les concepts de base d'un modèle multidimensionnel que la spécification de règles permettant de pondérer les composants d'un schéma multidimensionnel. En section 3, nous abordons la prise en compte de cette personnalisation lors des manipulations décisionnelles. La section 4 permet de mettre en évidence l'implantation de notre solution dans un contexte R-OLAP (Relational - On Line Analytical Processing).

### 2. Modélisation de Constellations Personnalisées

Notre modèle (Ravat, *et al.*, 2007) repose sur une représentation multidimensionnelle des données décrites au sein d'une *constellation* de *faits* et de *dimensions* munies de *hiérarchies* multiples.

## 2.1. Constellation

Une constellation, généralisation du modèle en étoile (Kimball, 1996), regroupe un ensemble de faits associés à des dimensions partagées et/ou spécifiques.

**Définition (constellation).** Une *constellation* C est définie par ($N^C$, $F^C$, $D^C$, $Star^C$) où $N^C$ est le nom de la constellation, $F^C = \{F_1, F_2,..., F_p\}$ est un ensemble de faits, $D^C = \{D_1, D_2,..., D_q\}$ est un ensemble de dimensions, et, $Star^C : F^C \rightarrow 2^{D^C}$ est une fonction associant les faits aux dimensions.

Une dimension modélise un axe d'analyse. Elle est caractérisée par des attributs organisés au sein d'une ou plusieurs hiérarchies.

**Définition (dimension).** Une *dimension* D est définie par ($N^D$, $A^D$, $H^D$, $I^D$) où $N^D$ est le nom de la dimension, $A^D = \{a_1, a_2,..., a_u\}$ est un ensemble d'attributs, $H^D = \{h^D_1, h^D_2,..., h^D_v\}$ est un ensemble de hiérarchies, et, $I^D = \{I^D_1, I^D_2,...\}$ est l'ensemble des instances de D.

Les hiérarchies organisent les attributs d'une dimension, appelés paramètres, du niveau de granularité d'analyse le plus fin, vers le niveau de granularité d'analyse le plus élevé. Ainsi une hiérarchie identifie les niveaux de granularité auxquels peuvent être manipulés les indicateurs d'analyse en spécifiant les chemins de navigation valides sur un axe d'analyse.

**Définition (hiérarchie).** Une *hiérarchie* $h^D_i$ est un chemin élémentaire acyclique partant du paramètre de plus faible granularité vers le paramètre de plus forte granularité. Elle est définie par ($N^{hD}_i$, $Param^{hD}_i$, $Suppl^{hD}_i$) où $N^{hD}_i$ est le nom de la hiérarchie, $Param^{hD}_i : P \rightarrow P$ ($P \subseteq A^D$) est une fonction décrivant la hiérarchie des attributs, appelés *paramètres* de la hiérarchie, $Suppl^{hD}_i : P \rightarrow 2^{(AD-P)}$ est une fonction spécifiant les *attributs faibles* qui complètent la sémantique des paramètres (un ensemble, éventuellement vide, d'attributs faibles est associé à un paramètre).

Un fait représente un ensemble d'indicateurs relatifs à un sujet d'analyse. Il est modélisé au travers de mesures représentant ces indicateurs.

**Définition (fait).** Un *fait* F est défini par le quadruplet ($N^F$, $M^F$, $I^F$, $IStar^F$) où $N^F$ est le nom du fait, $M^F = \{f_1(m_1), f_2(m_2),..., f_w(m_w)\}$ est un ensemble de *mesures* (ou indicateurs) agrégées selon une fonction $f_i \in \{AVG, SUM, MAX, MIN, COUNT,...\}$, $I^F = \{I^F_1, I^F_2,...\}$ est l'ensemble des instances de F, et, $IStar^F$ est une fonction associant les instances de $I^F$ aux instances des dimensions liées au fait.

**Exemple.** L'étude de cas illustrant les propositions de l'article concerne une société de vente par correspondance qui souhaite analyser les achats de produits ainsi que les ventes réalisées auprès de ses clients en France. Pour ce faire, nous proposons une constellation contenant deux faits et trois dimensions. Pour représenter simplement le schéma d'une constellation dans la figure 1, nous adoptons des notations graphiques proches de (Golfarelli *et al.* 1998).

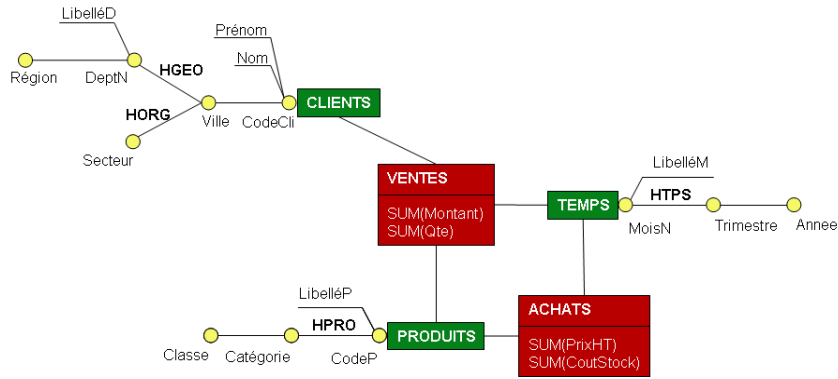

**Figure 1.** *Représentation graphique d'une constellation.*

## 2.2. Personnalisation de la constellation

Notre objectif est d'offrir les moyens à un utilisateur de paramétrer le schéma de la constellation de sorte à personnaliser son exploitation. Plus précisément, la personnalisation de la constellation s'opère au niveau des attributs de celle-ci, à savoir les paramètres, les attributs faibles et les mesures. En fonction des paramétrages effectués, les différents attributs sont alors utilisés ou non de manière prioritaire par le système lors de la manipulation de la constellation par l'utilisateur.

Nous proposons deux approches pour personnaliser une constellation.

– L'approche, dite « naïve » consiste à associer un poids aux attributs, fixant ainsi un ordre de priorité d'affichage à ceux-ci. Le système décide de l'utilisation d'un attribut en fonction de l'importance qui lui a été donnée.
– L'approche « avancée » consiste à définir un ordre de priorité d'un attribut en fonction du contexte de son utilisation, c'est-à-dire en fonction des opérations de manipulation et des données manipulées. Cette approche repose sur un paramétrage ECA (Widom, Ceri, 1996) de la constellation.

2.2.1. Personnalisation « naïve »

Une première approche consiste à associer un poids à chaque attribut $a_i$ de la constellation. Ce poids, noté $w_i$, modélise l'importance que l'utilisateur souhaite associer à $a_i$ ; afin de faciliter son exploitation ; chaque poids est normalisé $0 \leq w_i \leq 1$.

Lors de l'exploitation des données personnalisées, le système décisionnel affiche uniquement les données relatives aux attributs ayant un poids supérieur à un seuil fixé. Cette solution offre une grande flexibilité. Par défaut, nous avons fixé ce seuil à 1. Néanmoins, les autres attributs dont le poids est inférieur au seuil fixé restent accessibles lors des analyses décisionnelles. Les utilisateurs doivent le préciser explicitement. Ainsi, contrairement aux systèmes traditionnels où le paramètre de granularité maximale est initialement affiché et le décideur doit effectuer des opérations de forage pour afficher les autres paramètres, notre personnalisation affiche directement les paramètres possédant les poids les plus élevés.

**Exemple.** Considérons qu'un décideur souhaite construire un tableau de bord contenant les montants des achats trimestriels et annuels. Pour répondre à cet objectif, cet utilisateur doit affecter le poids 1 aux paramètres Année et trimestre de la dimension TEMPS.

L'atout d'une telle approche est la simplicité de mise en œuvre et d'exploitation par l'utilisateur. Néanmoins, elle souffre de certaines restrictions pouvant s'avérer pénalisantes dans le contexte d'une constellation. En effet, l'importance de certains paramètres peut varier suivant le **contexte d'utilisation**. Par exemple, un utilisateur peut exiger que l'analyse des ventes soit réalisée prioritairement de manière mensuelle tandis que celle concernant les achats peut nécessiter simplement une granularité trimestrielle. Ainsi, les poids associés aux paramètres MoisN et Trimestre sur la dimension TEMPS sont variables en fonction du contexte d'utilisation, à savoir en fonction des faits VENTES ou ACHATS.

2.2.2. Personnalisation « avancée »

Cette seconde approche consiste à personnaliser la constellation en intégrant le contexte d'utilisation de ses attributs. Pour cela, nous proposons un langage de configuration qui repose sur un langage de type ECA (Evènement – Condition – Action). La commande de définition d'une règle de configuration de la constellation repose sur la syntaxe la suivante :

```
CREATE RULE <nom_règle> ON { <N^D> | <N^{hD_i}> | <N^F> }
 WHEN <manipulation>
[IF <condition>]
 THEN <action>;
```

Les règles peuvent être associées soit à un fait ($N^F$), soit à une dimension ($N^D$), soit à une hiérarchie ($N^{hD_i}$).

`<manipulation>` : Cette expression détermine le contexte de la manipulation déclenchant la règle. Nous définissons ce contexte par rapport aux opérations de manipulation qui sont appliquées sur les composants de la constellation.

– **DISPLAYED**
– **ROTATED** [**FROM** $N^{Dold}$] [**TO** $N^{Dnew}$],
– **DRILLED-DOWN** [**ON** $N^{Dcurrent}$ [**TO** $p_{min}$] [**ACCORDING TO** $N^{HDcurrent}$]]
– **ROLLED-UP** [**ON** $N^{Dcurrent}$ [**TO** $p_{max}$] [**ACCORDING TO** $N^{HDcurrent}$]]

`<condition>` : Les règles peuvent être mises en place de manière conditionnelle. Une condition référence l'état courant de la constellation. Nous introduisons une fonction `current(E) : boolean` permettant de déterminer si un élément E de la constellation est en cours de manipulation ; $E \in \{N^D, N^D.N^{hD}_i, N^D[.N^{hD}_i].p_k, N^F, N^F.f_i, N^F[.f_i].m_k\}$.

`<action>` : Les actions s'appliquent sur les éléments constitutifs (faits, mesures, dimensions,...) de la constellation. Nous proposons une procédure prédéfinie `priority(E, `$w_i$`)` permettant d'associer contextuellement un poids à chaque attribut (mesure, paramètre, attribut faible). L'élément $E \in \{N^D, N^D.N^{hD}_i, N^D[.N^{hD}_i].p_k, N^F, N^F.f_i, N^F[.f_i].m_k\}$ ; lorsqu'il s'agit d'une dimension $N^D$, le poids $w_i$ est affecté à tous les attributs de la dimension, lorsqu'il s'agit d'une hiérarchie $N^D.N^{hD}_i$, le poids $w_i$ est affecté à tous les attributs de la hiérarchie… De manière analogue aux conditions, dans notre contexte de mise en œuvre R-OLAP, l'utilisateur peut définir ses propres actions sous forme de procédures stockées.

**Exemple 1.** Cet exemple présente la personnalisation de la dimension `Temps` suivant laquelle l'utilisateur modifie les priorités des attributs affichés lors de la manipulation de la dimension (`displayed`).

```
CREATE RULE display_temps_ventes ON Temps
WHEN displayed
THEN priority(Temps.HTPS.Année, 1),
     priority(Temps.HTPS.Trimestre, 1),
     priority(Temps.HTPS.MoisN, 0),
     priority(Temps.HTPS.LibelléM, 1);
```

Cette règle indique que les paramètres `Année` et `Trimestre` sont prioritaires tandis que le paramètre `MoisN` ne l'est pas. Il faut noter que l'attribut faible `LibelléM` du paramètre `MoisN` est spécifié prioritaire avec un poids comparable aux paramètres `Année` et `Trimestre`. Si l'affichage des mesures du fait « VENTES » est demandé en fonction de la dimension `Temps`, le système affiche automatiquement l'ensemble des attributs ayant le poids maximum ; il s'agit en l'occurrence dans cet exemple des paramètres `Année`, `Trimestre` et `LibelléM`.

**Exemple 2.** Dans l'exemple précédent, les priorités d'affichage des attributs sont fixées indépendamment de tout contexte. Ici, l'exemple personnalise l'affichage de la dimension `Temps` en fonction du contexte de son utilisation. La règle suivante précise que dans le contexte de l'analyse des ventes (clause `IF`) seront utilisés prioritairement les attributs `Année` et `MoisN`. Les trimestres seront affichés que si l'utilisateur exprime explicitement ce besoin. En outre, remarquons que cette personnalisation est valable dans le cadre d'opérations d'affichage et de rotation.

```
CREATE RULE display_temps_ventes ON Temps
WHEN displayed OR rotated
IF current(Ventes)
THEN priority(Temps.HTPS.Année, 1),
     priority(Temps.HTPS.Trimestre, 0),
     priority(Temps.HTPS.MoisN, 1);
```

**Exemple 3.** Ce troisième exemple, spécifie d'autres priorités lorsque le contexte d'utilisation de la dimension Temps concerne les achats (au lieu des ventes). Tandis que les ventes sont analysées de manière mensuelle en priorité, les analyses des achats sont réalisées trimestriellement en priorité.

```
CREATE RULE display_temps_achats ON Temps
WHEN displayed
IF current(Achats)
THEN priority(Temps.HTPS.Année, 1),
     priority(Temps.HTPS.Trimestre, 1),
     priority(Temps.HTPS.MoisN, 0);
```

La combinaison de la clause WHEN et de la clause IF permet de distinguer deux types de personnalisation contextuelle : soit le **contexte de manipulation**, soit le **contexte d'utilisation**. Le contexte de manipulation permet de spécifier l'opération suivant laquelle les priorités sont fixées tandis que le contexte d'utilisation spécifie l'état courant de la constellation suivant lequel les priorités doivent être fixées. Ces deux contextes peuvent être utilisés simultanément au sein d'une même règle.

## 3. Manipulations OLAP Personnalisées

Lors des manipulations opérées par les utilisateurs, la structure de visualisation, appelée table multidimensionnelle, centre l'analyse sur un fait courant en fonction de deux dimensions courantes hiérarchisées. Les manipulations consistent à appliquer sur la table multidimensionnelle un ensemble d'opérations OLAP pour transformer les données visualisées provenant de la constellation (Ravat, *et al.* 2006).

### 3.1. Table Multidimensionnelle : TM

**Définition (table multidimensionnelle).** Une table multidimensionnelle T est définie par (S, L, C, R) où S = (F, {$f_1(m_1)$, $f_2(m_2)$,…}) représente le sujet d'analyse relatif au fait F et ses mesures observées {$m_1$, $m_2$,…} agrégées à l'aide de fonctions $f_1, f_2,…$, L = (DL, HL, <All, $p^{DL}_1$, …, $p^{DL}_v$>) représente l'axe d'analyse en ligne de T au travers d'une dimension courante DL, d'une hiérarchie courante HL et d'une liste ordonnée de paramètres affichés <$p^{DL}_1$, $p^{DL}_2$,…>, C = (DC, HC, <All, $p^{DC}_1$,…, $p^{DC}_w$>) représente l'axe d'analyse en colonne de la table T au travers d'une dimension courante DC, d'une hiérarchie courante HC et d'une liste ordonnée de paramètres affichés <$p^{DC}_1$, $p^{DC}_2$,…>, et, R = $pred_1 \wedge pred_2 \wedge …$ est le prédicat de restriction composé d'une conjonction de prédicats normalisés portant sur les dimensions et/ou F.

**Exemple.** La TM décrite à la figure 2 visualise la somme des montants des ventes par région des clients et par classe de produits. Elle est définie comme suit :

– le sujet d'analyse courant S = (VENTES, {*SUM*(montant)}),

- la dimension courante en ligne L = (CLIENTS, HGEO, <Région>),
- la dimension courante en colonne C = (PRODUITS, HPRO, <Classe>),
- le prédicat de restriction est vide car toutes les données disponibles dans la constellation sont utilisées

| VENTES | | PRODUITS.HPRO | | | |
|---|---|---|---|---|---|
| | SUM(Montant) | Classe | Technologique | Habilement | Mobilier |
| CLIENTS.HGEO | Région | | | | |
| | Midi-Pyrénées | | 2 000,00 € | 3 500,00 € | 1 500,00 € |
| | Aquitaine | | 1 800,00 € | 3 000,00 € | 2 000,00 € |
| | Bretagne | | 1 600,00 € | 3 200,00 € | 1 900,00 € |

**Figure 2.** *Exemple de table multidimensionnelle.*

### 3.2. Opérations de manipulations étendues

La personnalisation proposée consiste à associer aux attributs des poids reflétant l'intérêt prioritaire que porte l'utilisateur aux attributs d'une constellation. Ces priorités sont exploitées par le système lors des manipulations utilisateurs. Pour ce faire, nous proposons d'étendre certains opérateurs supportant les manipulations OLAP (Ravat, *et al.*, 2006). Dans cet article, nous limitons notre étude aux principaux opérateurs : le constructeur de table multidimensionnelle (DISPLAY), la rotation (ROTATE) et les forages (DRILL-DOWN et ROLL-UP). En effet, ces opérations induisent dans leur fonctionnement classique l'ajout ou la suppression implicite d'attributs dans une table multidimensionnelle. La personnalisation que nous avons introduite vient transformer leur fonctionnement traditionnel afin d'adapter leur fonctionnement aux besoins exprimés par l'utilisateur.

Le tableau 1 donne la représentation algébrique des opérateurs étendus.

| Opérateurs | Représentations algébriques |
|---|---|
| Constructeur | `DISPLAY(F,{`$f_1(m_1)$`,…},DL,HL,DC,HC [, seuil])=`$T_{RES}$ |
| Rotation | `DROTATE(`$T_{SRC}$`,`$D_{old}$`,`$D_{new}$`,`$H^{Dnew}_k$` [, seuil])=`$T_{RES}$ |
| Forage vers le bas | `DRILLDOWN(`$T_{SRC}$`,D,`$Att_{inf}$` [, seuil])=`$T_{RES}$ |
| Forage vers le haut | `ROLLUP(`$T_{SRC}$`,D,`$Att_{sup}$` [, seuil])=`$T_{RES}$ |

**Tableau 1.** *Opérateurs étendus.*

DISPLAY permet d'extraire les composants d'une BDM pour créer une première TM tandis que les autres opérateurs respectent la propriété de fermeture en reposant sur une TM en entrée et en produisant une TM résultat en sortie. Cette propriété offre ainsi un cadre permettant d'élaborer des requêtes complexes par combinaisons d'opérations élémentaires.

L'extension consiste en l'ajout d'un seuil de priorité avec lequel les poids des attributs sont comparés. Ce seuil est optionnel ; lorsque le seuil n'est pas spécifié, l'opérateur se place dans son fonctionnement traditionnel où les poids sont ignorés. Lorsque le seuil est exprimé, tout attribut dont le poids est supérieur au seuil se voit pris en compte dans l'affichage résultat de l'opération.

3.2.1. Constructeur

Dans son fonctionnement classique, DISPLAY spécifie le fait et ses mesures ainsi que les dimensions et hiérarchies en ligne et colonne. Pour chaque dimension, le système affiche le paramètre de granularité maximale de la hiérarchie spécifiée.

**Exemple.** `DISPLAY(VENTES, {`*SUM*`(Montant)}, CLIENTS, HGEO, PRODUITS, HPRO)=T`$_{RES1}$ produit la table multidimensionnelle de la figure 2.

Dans un fonctionnement personnalisé, les paramètres des dimensions courantes peuvent être complétés et/ou remplacés par d'autres paramètres en fonction des règles de personnalisation.

**Exemple.** Supposons que l'analyse des ventes selon l'axe clients s'effectue de manière prioritaire en fonction du département puis de manière secondaire en fonction des régions puis enfin en fonction de la ville et du code client. Cet ordonnancement des priorités peut s'exprimer au travers de la règle suivante :

```
CREATE RULE display_clients ON Clients
WHEN displayed
IF current(HGEO)
THEN priority(Produits.HGEO.CodeCli, 0.4),
     priority(Produits.HGEO.Ville,   0.4),
     priority(Produits.HGEO.DeptN,   0.8),
     priority(Produits.HGEO.Région,  0.6);
```

L'opération `DISPLAY(VENTES, {`*SUM*`(Montant)}, CLIENTS, HGEO, PRODUITS, HPRO, 0.5)=T`$_{RES2}$ construit une TM personnalisée où le paramètre `Région` normalement affiché est complété par le paramètre `DeptN (cf. figure 3)`.

| VENTES | | | PRODUITS.HPRO | | | |
|---|---|---|---|---|---|---|
| | SUM(Montant) | | Classe | Technologique | Habilement | Mobilier |
| CLIENTS.HGEO | Région | DeptN | | | | |
| | Midi-Pyrénées | 31 | | 1 200,00 € | 2 000,00 € | 1 000,00 € |
| | | 81 | | 800,00 € | 1 500,00 € | 500,00 € |
| | Aquitaine | 33 | | 1 800,00 € | 3 000,00 € | 2 000,00 € |
| | Bretagne | 22 | | 800,00 € | 2 000,00 € | 1 000,00 € |
| | | 29 | | 800,00 € | 1 200,00 € | 900,00 € |

**Figure 3.** *Table multidimensionnelle personnalisée.*

Notons que cette version étendue de l'opération DISPLAY n'affiche pas nécessairement le paramètre de plus haute granularité (contrairement à son fonctionnement classique) mais permet d'afficher automatiquement tous les éléments dont le poids est supérieur au seuil. Ainsi, cette version étendue permet un affichage personnalisé tout en diminuant le nombre de commandes à exécuter (notamment les forages vers le bas).

3.2.2. Rotation

Le fonctionnement classique de la rotation consiste à changer une dimension d'une TM par une nouvelle dimension. Par défaut, le système se positionne sur le paramètre de granularité maximale de la hiérarchie désirée de la nouvelle dimension.

**Exemple.** ROTATE(T$_{RES1}$, PRODUITS, TEMPS, HTPS)=T$_{RES3}$ produit une nouvelle TM où la dimension en ligne PRODUITS est remplacée par TEMPS. Le paramètre Année est automatiquement utilisé comme paramètre courant dans la TM résultat. Les montants visualisés sont recalculés en fonction de ces nouveaux paramètres.

Dans le cadre d'une rotation personnalisée, de la même manière que pour le constructeur de table, tous les attributs dont le poids est supérieur à la valeur du seuil considéré dans la commande sont automatiquement affichés.

**Exemple.** Considérons la personnalisation de la dimension Temps définie à la section 2.2.2. Les règles étant définies par rapport au seul contexte de déclenchement DISPLAYED, il convient de compléter la clause WHERE par l'expression 'OR Rotated To Temps' pour les rendre actives lorsque qu'une rotation remplace une dimension par la dimension Temps. L'opération de rotation ROTATE(T$_{RES1}$, PRODUITS, TEMPS, HTPS, 0.5) produit des résultats suivants :

– Selon la personnalisation de l'exemple 1 section 2.2.2, les paramètres Année, Trimestre et LibelléM sont affichés dans la table résultat.
– Selon la personnalisation de l'exemple 2 section 2.2.2, les paramètres Année et MoisN sont affichés.

3.2.3. Forages vers le bas et vers le haut

L'opération de forage vers le bas consiste de manière classique à compléter les paramètres courants d'une dimension affichée par un paramètre de granularité inférieure sur la hiérarchie courante. A l'inverse, l'opération de forage vers le haut supprime les paramètres de plus basses granularités en se repositionnant sur un paramètre de plus grande granularité.

**Exemple.** A l'issue de l'opération de rotation précédente, dans son fonctionnement classique, la table multidimensionnelle T$_{RES3}$ est produite avec pour seul paramètre le paramètre Année. Une opération de forage vers le bas peut permettre à l'utilisateur d'affiner l'analyse en ajoutant le paramètre MoisN : DRILLDOWN(T$_{RES3}$, TEMPS, MoisN)=T$_{RES4}$. Notons que le paramètre Trimestre n'est pas affiché (pour l'obtenir l'utilisateur aurait dû exprimer DRILLDOWN(DRILLDOWN(T$_{RES3}$, TEMPS, Trimestre), TEMPS, MoisN)).

Dans le cas de forages personnalisés, les poids sont utilisés par le système pour déterminer les paramètres intermédiaires affichés. Notons que le paramètre explicitement défini dans l'opération est affiché indépendamment de son poids.

**Exemple.** Nous reconsidérons la table multidimensionnelle $T_{RES3}$ et supposons qu'un utilisateur définisse la règle suivante :

```
CREATE RULE display_temps_ventes ON Temps
WHEN displayed
IF current(Ventes)
THEN priority(Temps.HTPS.Année, 0.7),
     priority(Temps.HTPS.Trimestre, 1),
     priority(Temps.HTPS.LibelléM, 0.7),
     priority(Temps.HTPS.MoisN, 0.5);
```

L'opération de forage vers le bas personnalisée DRILLDOWN($T_{RES3}$, TEMPS, MoisN, 0.6) permet à l'utilisateur d'affiner l'analyse en ajoutant le paramètre MoisN, mais également le paramètre intermédiaire Trimestre. Le paramètre MoisN, demandé explicitement dans l'opération, est affiché et complété par l'attribut faible LibelléM qui n'ajoute aucune granularité supplémentaire, mais qui introduit une information sémantique additionnelle à celle de MoisN.

Un forage vers le haut DRILLDOWN($T_{RES3}$, TEMPS, Trimestre, 0.8) permet d'obtenir sur la dimension Temps uniquement le paramètre Trimestre tandis que le paramètre Année, non qualifié par le poids de 0.7 au regard du seuil de 0.8 disparaît.

## 4. Mise en œuvre en R-OLAP

### 4.1. Architecture globale

Pour valider notre proposition, nous développons un prototype reposant sur l'architecture décrite dans la figure 4. Il repose essentiellement sur deux modules :

– un module « **pré-compilateur** » chargé d'analyser (lexicalement et syntaxiquement) les règles saisies par l'utilisateur, puis de les traduire dans la BD R-OLAP sous forme de requêtes de mises à jour dans des tables systèmes,
– un module « **analyseur de requêtes** » chargé d'analyser les requêtes de manipulation et de construire les tables multidimensionnelles résultats en fonction des seuils exprimés et des poids affectés.

La constellation de faits et de dimensions est implantée sous forme de relations non représentées dans la figure suivante.. Pour simplifier nous limitons notre présentation aux relations principales ; nous n'abordons pas la problématique des relations auxiliaires dérivées appelées pré-agrégats et implantés au travers de vues matérialisées.

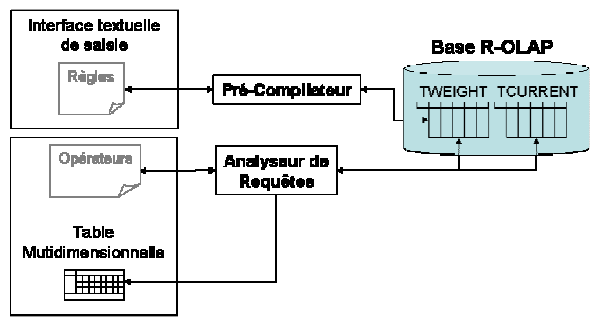

**Figure 4.** *Architecture générale du système décisionnel personnalisable.*

### 4.2. Accès personnalisé aux relations principales

Les tables systèmes, TCURRENT et TWEIGHT, sont chargées respectivement de « tracer » les éléments en cours d'interrogation dans la TM et de connaître les poids des attributs de la constellation. Par mesure de simplification, nous ne présentons que les colonnes impliquées par les exemples traités dans l'article.

#### 4.2.1. Pré-compilateur

**Exemple.** Pour la TM décrite en figure 2, la table système TCURRENT contient les enregistrements représentant le contexte d'utilisation, c'est-à-dire les informations concernant les éléments intervenant dans l'affichage de la TM courante.

| TCURRENT | | |
|---|---|---|
| ELEMENT | TYPE_ELT | POSITION |
| VENTES | F | CEL |
| MONTANT | M | CEL |
| CLIENTS | D | ROW |
| HGEO | H | ROW |
| Région | P | ROW |
| PRODUITS | D | COL |
| HPRO | H | COL |
| Classe | P | COL |

**Figure 5.** *Exemple d'enregistrements de la table système TCURRENT.*

La prise en compte des règles repose sur un pré-compilateur qui effectue une mise à jour de la table système TWEIGHT ainsi que des déclencheurs chargés de maintenir la table cohérente au regard des contextes d'utilisation.

**Exemple.** La figure 6 présente un exemple de TWEIGHT en fonction de règle de l'exemple 1 section 2.2.2. Chaque ligne de la table représente le poids d'un attribut de la constellation. La colonne ELEMENT contient la dimension ou le fait, la colonne HIERARCHY contient la hiérarchie (pour un fait la valeur est NULL). Cette colonne permet d'affecter des poids différents aux attributs partagés entre plusieurs hiérarchies. La colonne OPERATION indique l'opération suivant laquelle les poids doivent être pris en compte. Notez que pour les opérations de rotation et de forages, d'autres colonnes non présentées participent au schéma cette table (par exemple pour connaître les valeurs FROM et TO du ROTATED).

| TWEIGHT | | | | |
|---|---|---|---|---|
| OPERATION | ELEMENT | HIERARCHY | ATTRIBUTE | WEIGHT |
| DISPLAYED | TEMPS | HTPS | Année | 1 |
| DISPLAYED | TEMPS | HTPS | Trimestre | 1 |
| DISPLAYED | TEMPS | HTPS | LibelléM | 1 |
| DISPLAYED | TEMPS | HTPS | MoisN | 0 |
| ROTATED | TEMPS | HTPS | Année | 1 |
| ROTATED | TEMPS | HTPS | Trimestre | 1 |
| ROTATED | TEMPS | HTPS | LibelléM | 1 |
| ROTATED | TEMPS | HTPS | MoisN | 0 |

**Figure 6.** *Exemple d'enregistrements de TWEIGHT.*

4.2.2. Analyseur de requêtes

Chaque manipulation de l'utilisateur est transcrite sous forme d'opérations algébriques étendues (cf. section 3.2). L'analyseur réalise les actions suivantes :

– détermination des poids des attributs en fonction de TWEIGHT, le système active des déclencheurs pour ajuster les poids en fonction du contexte d'utilisation stocké dans TCURRENT,
– calcul des données R-OLAP à analyser en fonction de la requête,
– construction de la table multidimensionnelle.

**Exemple.** La requête ROTATE(T$_{RES1}$, PRODUITS, TEMPS, HTPS) est utilisée par l'analyseur de requêtes. Le système calcule alors l'axe TEMPS dans la nouvelle table multidimensionnelle en utilisant tous les attributs qualifiés par leur poids. La table système TWEIGHT intervient pour réajuster dynamiquement les poids en fonction du contexte de manipulation.

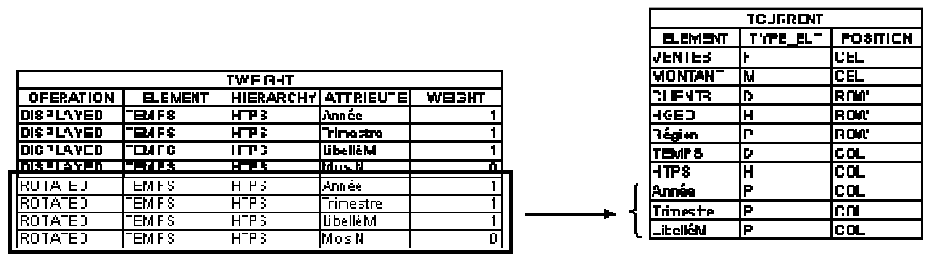

**Figure 7.** *Détermination de l'affichage personnalisé.*

## 5. Conclusion

Notre proposition se situe dans le cadre des systèmes décisionnels reposant sur une modélisation multidimensionnelle des données. Plus précisément, nous avons défini des concepts et des mécanismes permettant de personnaliser l'affichage et la manipulation de données multidimensionnelles.

Dans un premier temps, nous avons défini un modèle multi-sujets d'analyses selon différents axes d'analyse multi-hiérarchisés. Ce modèle, que nous appelons constellation, est complété par un langage de définition de règles ECA. Ces règles permettent de personnaliser une constellation en affectant un poids aux attributs. La valeur de ce poids, comprise entre 0 et un 1, permet de spécifier un affichage adapté en fonction du contexte et notamment pour les paramètres partagés entre plusieurs hiérarchies d'une même dimension ou les dimensions partagées entre plusieurs faits.

Cette personnalisation induit également un fonctionnement spécifique des opérations classiquement utilisées lors d'analyses décisionnelles. Le constructeur de table multidimensionnelle permet d'afficher en une seule commande l'ensemble des attributs dont le poids est supérieur à un seuil déterminé par le décideur. Les rotations, changements d'axes d'analyse au sein d'une table multidimensionnelle, peuvent également s'effectuer en fonction d'un seuil et ainsi afficher tous les attributs de la nouvelle dimension dont le poids est supérieur au seuil. Les opérations de forages reposent également sur le même principe.

Notre proposition conceptuelle a été validée par le développement d'un prototype permettant de spécifier et d'exploiter des règles de personnalisation ECA. Il repose sur des développements réalisés en Java (interfaces), JavaCC (analyseur lexical et syntaxique), et Oracle (implantations R-OLAP).

Les extensions à ces travaux sont nombreuses. Dans un premier temps, nous pouvons proposer une personnalisation encore plus fine d'un schéma en constellation. Cette personnalisation consiste non seulement à spécifier des attributs mais également des ensembles de valeurs (comme par exemple, travailler sur les 3 dernières années). Cette personnalisation plus fine permet également d'affecter un poids commun à l'ensemble des attributs d'une dimension voire d'une hiérarchie associé à un contexte d'utilisation particulier tel qu'une opération de construction, de forage ou de rotation. Dans un second temps, notre objectif est d'aboutir à un système adaptatif permettant de présenter l'ensemble des données essentielles au travers de rapports prédéfinis aux décideurs, limitant ainsi les différentes opérations couramment effectuées par ceux-ci. Le système adaptatif pourrait engendrer automatiquement des règles de personnalisation afin d'ajuster le système décisionnel aux usages de chaque décideur.

## 6. Bibliographie